\documentstyle[mnextra]{mn}
\input{epsf}
\seceqnum           
\def\etal{{\rm et al.} }
\def\kms  {{\rm km \, s^{-1}}}
\def\skms  {\,{\rm km \, s^{-1}}}
\def\Mpc  {{\it h}^{-1}\,{\rm Mpc}}
\def\sMpc  {\,{\it h}^{-1}\,{\rm Mpc}}

\def\sMsol {\,{\it h}^{-1}\,{\rm M_\odot}}
\def\d    {{\rm d}}
\def\ie {{\rm i.e. }}
\def\eg {{\rm e.g. }}
\def\log {{\rm log}}

\def\lsim{\mathrel{\hbox{\rlap{\hbox{\lower4pt\hbox{$\sim$}}}\hbox{$<$}}}}
\def\gsim{\mathrel{\hbox{\rlap{\hbox{\lower4pt\hbox{$\sim$}}}\hbox{$>$}}}}

\def\bj{b_{\rm J}}
\def\bjlim{b_{\rm J}^{\rm lim}}
\def\bjbright{{b_{\rm J}^{\rm bright}}}
\def\btheta{\theta}

\begin{document}
\title[2dFGRS galaxy groups]
{Galaxy groups in the 2dFGRS: the group-finding algorithm and the 2PIGG
catalogue}
\author[V.R. Eke et al. (the 2dFGRS Team)]
{\parbox[t]\textwidth{
V.R. Eke$^{1}$,
Carlton M. Baugh$^{1}$,
Shaun Cole$^1$,
Carlos S. Frenk$^1$,
Peder Norberg$^{2}$,
John A. Peacock$^{3}$,
Ivan K. Baldry$^{4}$,
Joss Bland-Hawthorn$^5$,
Terry Bridges$^5$,
Russell Cannon$^5$,
Matthew Colless$^6$,
Chris Collins$^7$,
Warrick Couch$^8$,
Gavin Dalton$^{9,10}$,
Roberto de Propris$^8$,
Simon P. Driver$^6$,
George Efstathiou$^{11}$,
Richard S. Ellis$^{12}$,
Karl Glazebrook$^4$,
Carole A. Jackson$^6$,
Ofer Lahav$^{11}$,
Ian Lewis$^9$,
Stuart Lumsden$^{13}$,
Steve J. Maddox$^{14}$, 
Darren Madgwick$^{15}$,
Bruce A.\ Peterson$^6$,
Will Sutherland$^{10}$,
Keith Taylor$^{12}$ (the 2dFGRS Team)}
\vspace*{6pt} \\
{$^1$Department of Physics, University of Durham, South Road,
    Durham DH1 3LE, UK} \\
{$^2$ETHZ Institut f\"ur Astronomie, HPF G3.1, ETH H\"onggerberg, CH-8093
       Z\"urich, Switzerland} \\
{$^3$Institute for Astronomy, University of Edinburgh, Royal 
       Observatory, Blackford Hill, Edinburgh EH9 3HJ, UK}\\
{$^4$Department of Physics \& Astronomy, Johns Hopkins University,
       Baltimore, MD 21218-2686, USA} \\
{$^5$Anglo-Australian Observatory, P.O.\ Box 296, Epping, NSW 2121,
    Australia}\\
{$^6$Research School of Astronomy and Astrophysics,
 The Australian National University, Canberra, ACT 2611, Australia}\\
{$^7$Astrophysics Research Institute, Liverpool John Moores University,
    Twelve Quays House, Birkenhead, L14 1LD, UK} \\
{$^8$Department of Astrophysics, University of New South Wales, Sydney,
    NSW 2052, Australia} \\
{$^9$Department of Physics, University of Oxford, Keble Road,
    Oxford OX1 3RH, UK} \\
{$^{10}$Rutherford Appleton Laboratory, Chilton, Didcot, OX11 0QX} \\
{$^{11}$Institute of Astronomy, University of Cambridge, Madingley Road,
    Cambridge CB3 0HA, UK} \\
{$^{12}$Department of Astronomy, California Institute of Technology,
    Pasadena, CA 91125, USA} \\
{$^{13}$Department of Physics, University of Leeds, Woodhouse Lane,
       Leeds, LS2 9JT, UK} \\
{$^{14}$School of Physics \& Astronomy, University of Nottingham,
       Nottingham NG7 2RD, UK} \\
{$^{15}$Department of Astronomy, University of California, Berkeley, CA 92720, USA}\\ 
}

\maketitle

\begin{abstract}

The construction of a catalogue of galaxy groups from the 2-degree
Field Galaxy Redshift Survey (2dFGRS) is described. Groups are
identified by means of a friends-of-friends percolation algorithm
which has been thoroughly tested on mock versions of the 2dFGRS
generated from cosmological N-body simulations. The tests suggest that
the algorithm groups all galaxies that it should be grouping, with an
additional 40 per cent of interlopers.  About $55$ per cent of the
$\sim 190\,000$ galaxies considered are placed into groups containing at
least two members of which $\sim 29\,000$ are found. Of these, $\sim
7000$ contain at least four galaxies, and these groups have a median
redshift of $0.11$ and a median velocity dispersion of $260
\skms$. This 2dFGRS Percolation-Inferred Galaxy Group (2PIGG) catalogue
represents the largest available homogeneous sample of galaxy
groups. It is publicly available on the WWW.

\end{abstract}
\begin{keywords}
catalogues -- galaxies: clusters: general.
\end{keywords}

\section{Introduction}

Groups of galaxies are useful tracers of large-scale structure. They
provide sites in which to study the environmental dependence of galaxy
properties, the galactic content of dark matter haloes, the
small-scale clustering of galaxies, and the interaction between
galaxies and hot X-ray emitting intragroup gas. Thus, it is desirable
to have an extensive, homogeneous catalogue of groups of galaxies
representing the various bound systems in the local universe.

Many studies in the past relied upon the pioneering work of Abell
(1958) to provide a set of target galaxy clusters (see also Abell,
Corwin \& Olowin 1989, Lumsden \etal 1992 and Dalton \etal 1997 for
similar studies). However, because of
the lack of redshift information available when Abell was defining his
cluster catalogue, concerns have been raised over the completeness of
his sample and the impact that line-of-sight projections would have in
contaminating it (\eg Lucey 1983; Sutherland 1988; Frenk \etal 1990;
van Haarlem, Frenk \& White 1997). As a result of these worries, it
became fashionable to select galaxy cluster samples based upon cluster
X-ray emission (\eg Gioia \etal 1990; Romer 1995; Ebeling \etal
1996; B\"ohringer \etal 2001), 
this method being less prone to projection effects. This
strategy nevertheless brings its own complications, because X-ray
emission depends sensitively on the details of intracluster gas
physics.

The construction of galaxy redshift surveys over the past twenty years
has enabled a number of groups to pursue the optical route to
group-finding. Huchra \& Geller (1982), Geller \& Huchra (1983),
Nolthenius \& White (1987), Ramella, Geller \& Huchra (1989) and
Moore, Frenk \& White (1993) used subsets of the Center for
Astrophysics (CfA) redshift survey, containing a few thousand
galaxies, to investigate both the abundance and the internal
properties of samples of a few hundred galaxy groups. Maia, da Costa
\& Latham (1989) extended the database using the Southern Sky Redshift
Survey of a further $\sim 1500$ galaxies and identified a sample of
$87$ groups containing at least two members. All-sky galaxy samples of
$\sim 2400$, $4000$ and $6000$ galaxies were used for group-finding by
Tully (1987), Gourgoulhon, Chamaraux \& Fouqu\'e (1992) and Garcia
(1993) respectively. Deeper surveys of small patches of sky have also been 
used to find groups by Ramella \etal (1999, $\sim 3000$ galaxies) and Tucker 
\etal (2000, $\sim 24\,000$ galaxies).
Giuricin \etal (2000), Merch\'an, Maia \& Lambas
(2000) and Ramella et al (2002) have performed analyses on catalogues
containing up to $\sim 20\,000$ galaxies, but the largest set of galaxy
groups so far is that provided by Merch\'an \& Zandivarez (2002). They
used the $\sim 60\,000$ galaxies in the contiguous Northern and Southern
Galactic Patches (NGP and SGP) in the `100k' public data release of
the 2dFGRS (Colless \etal 2001).  This work extends that sample to the
NGP and SGP regions in the complete 2dFGRS\footnote{The 2dFGRS data
release is described by Colless \etal (astro-ph/0306581) and the data can be
found on the WWW at http://www.mso.anu.edu.au/2dFGRS/.}. 
The entire survey contains
about 220\,000 galaxies, $\sim 190\,000$ of which are in the two
contiguous patches once the completeness cuts detailed below have been
applied.

As well as using more galaxies than were previously available, this study
also contains the results from rigorous tests of the group-finding
algorithm. These are facilitated by the construction of very
detailed mock 2dFGRSs, created using a combination of
dark matter simulations and semi-analytical galaxy formation models. 
Comparing the properties of the recovered galaxy groups to those of
the underlying parent dark matter haloes provides a robust
framework within which the parameters of the group-finding algorithm
can be chosen so as to apportion the galaxies optimally. Furthermore,
the ability to relate, in a quantitative fashion, the input mass and
galaxy distributions to the set of recovered galaxy groups, is of crucial
importance when trying to extract scientific information from the
group catalogue. This approach is identical in spirit to that adopted by 
Diaferio \etal (1999) when they studied the northern region of the CfA
redshift survey.

The choice of group-finding algorithm is described in
Section~\ref{sec:fof}. Details of the
mock catalogue construction and group-finder testing are given in
Sections~\ref{sec:mock} and ~\ref{sec:test}. Section~\ref{sec:2df}
contains the results when the group-finder is applied to the real 2dFGRS.

\section{The group-finder}\label{sec:fof}

\subsection{Choice of algorithm}\label{ssec:alg}

Given a set of galaxies with angular positions on the sky and
redshifts, the task of the group-finder is broadly to return sets of
galaxies that are most likely to represent the true bound structures
that are being traced by the observed galaxies. For some applications,
not missing any of the true group members will be particularly
desirable. For others, minimising the amount of contamination by
nearby, yet physically separate, objects will be the priority.  It is
inevitable that some distinct collapsed objects, situated near to each
other in real space and along a similar line-of-sight, will be spread
out by line-of-sight velocities to the extent that they overlap with
one another. Thus, some unavoidable contamination is to be
expected. The aim of the group-finder described here is to find a
compromise between the extremes of finding all true members and
minimising contamination, with a view to providing groups that have
velocity dispersions and projected sizes similar to those of their
parent dark matter haloes. Naturally, the efficiency of such a
group-finder can only be calibrated and tested when the properties of
the associated dark matter haloes are known. The use of realistic mock
catalogues is thus central to the entire group-finding procedure
because it directly affects the composition of the final group
catalogue through the choice of parameters for the group-finder.

To date, the job of finding groups in galaxy redshift surveys has
typically been assigned to a percolation algorithm (see Tully 1987,
Marinoni \etal 2002 and references within Trasarti-Battistoni 1998
for alternative approaches) that links together all
galaxies within a particular linking volume centred on each
galaxy. These friends-of-friends (FOF) algorithms are specified
by the shape and size of the linking volume and how it varies
throughout the survey. In order to produce galaxy groups corresponding to
a similar overdensity throughout the survey, the linking volume should
be scaled to take into account the varying number density of galaxies
that are detected as a function of redshift. Previous studies have not
all chosen the same scaling for the linking length with mean
observed galaxy number
density, $n$. At higher redshifts, flux-limited catalogues contain a
lower number density of galaxies, causing the mean intergalaxy
separation to increase.
The algorithm proposed by Huchra \& Geller (1982) scales both the
perpendicular (in the plane of the sky) and the parallel
(line-of-sight) linking lengths ($\ell_\perp$ and $\ell_{||}$) in proportion
to $n^{-1/3}$. Ramella, Geller \& Huchra (1989) scaled
both linking lengths in proportion to $n^{-1/2}$.
In contrast, Nolthenius \& White
(1987) and Moore, Frenk \& White (1993) chose to set $\ell_{||}$
to correspond to the typical size in redshift space of groups detected as
a function of redshift. The perpendicular linking lengths were scaled
in proportion to $n^{-1/2}$, this being how the mean projected
separation varies. Thus, in contrast with the other methods, the
aspect ratio of this linking volume is not independent of redshift.

In choosing how to scale the linking lengths, there is one primary
condition that one would like to satisfy. Namely, that for a
particular group of galaxies sampled at varying completeness, the
edges of the recovered group should be in similar places. If this is
achieved, then the inferred velocity dispersion and projected size and the
actual contamination should be independent of the sampling rate.
Scaling both $\ell_\perp$ and $\ell_{||}$ by $n^{-1/3}$ will lead to
groups of similar shape and overdensity being found throughout the
survey. Of course if the galaxy distribution is sampled very sparsely
then this scaling can lead to linking lengths that are large with
respect to the size of real gravitationally bound structures so,
depending upon the nature of the galaxy survey, it may be
desirable to put an upper limit on the size of the linking
volume. The maximum value of the perpendicular linking length is one
of the parameters of the algorithm used here.

\subsection{The linking volume}\label{ssec:link}

Having defined how the linking volume scales with mean observed galaxy
number density, the choice of its shape and size still remains.  The
shape of the linking volume should clearly be spherical in the case
where real galaxy distances are measured. However, with redshift space
distances, groups will appear elongated along the line-of-sight, and
one can appreciate that in order to find all of the group members,
$\ell_{||}>\ell_\perp$ would be helpful. An approximate estimate of the
amplitude of this stretching can be made by considering how rapidly
galaxies will be moving in a halo of mass $M$ and radius $r$. If the
circular velocity satisfies $v^2=GM/r$, the line-of-sight velocity
dispersion can be written as $\sigma^2 \approx v^2/2$ and the total
mass is $M=4/3 \pi r^3 \Delta_c \rho_c$, where $\rho_c$ is the
critical density and $\Delta_c$ defines the mean density relative to
critical within the halo, then
\begin{equation}
\frac{\sigma}{r} \approx \sqrt{\frac{2\pi G\Delta_c\rho_c}{3}}.
\label{lrat1}
\end{equation}
Setting $\Delta_c=150$ yields
\begin{equation}
\frac{\sigma}{r} \approx 600\skms/(\Mpc).
\label{lrat2}
\end{equation}
(Note that, according to the spherical `top-hat' model
$\Delta_c\approx 100,180$ for $\Omega_0=0.3,\Lambda_0=0.7$ and
$\Omega_0=1$ models respectively; Eke, Cole \& Frenk 1996) Thus, for
an object with virial radius $r$, the velocities produce a $1\sigma$
stretch along the line-of-sight of $\sim 6r$. The ratio of parallel to
perpendicular sizes is independent of $r$. 
This ignores the redshift
measurement errors which will add in an $r$-independent parallel
contribution of $\sigma_{\rm err}\sim 85\skms$ for the 2dFGRS (Colless
\etal 2001) with no corresponding perpendicular increase. Even at the
lowest redshifts this error term is smaller than the typical
line-of-sight linking length for the 2dFGRS.
Consequently, the shape
of the linking volume should be elongated, with a particular aspect
ratio given by $R_{\rm gal}=\ell_{||}/\ell_{\perp}$, and the appropriate
value for this ratio should be $\sim 12$ to enclose $2\sigma$ of the
galaxies along the line-of-sight.

In dark matter N-body simulations, friends-of-friends group-finders are often
applied in real space with
a linking length that is $b=0.2$ times the mean interparticle
separation in order to identify groups having overdensities that are
$\sim b^{-3}$ (Davis \etal 1985). 
This choice of linking length has been shown by Jenkins \etal (2001)
to yield a halo mass function that is independent of redshift and
$\Omega_0$, and thus provides a good definition of the underlying dark
matter haloes. Since redshift space, rather than
real space distances are available
in the 2dFGRS, it is unclear quite what value of
$b_{\rm gal}$, in terms of the mean intergalaxy separation, is appropriate to
reproduce the boundary of the groups as defined by a $b=0.2$ set of
dark matter groups. The parameter $b$ will be used to set the
overall size of the linking volume through
$\ell_\perp=b/n^{1/3}$.
The semi-analytical model of Cole \etal (2000) predicts that light is
typically more concentrated than mass in groups. Thus, a value of
$b_{\rm gal}$ smaller than $0.2$ is likely to be appropriate
for recovering the $b=0.2$ set of dark matter groups from a galaxy survey.

This discussion implies that the linking volume should be elongated
along the redshift direction, and provides rough expectations for both
the aspect ratio and overall size of the volume that will best recover
the underlying dark matter haloes. However, the precise shape of this
volume is yet to be specified. Both a cylinder and an ellipsoid could
satisfy the above requirements. In the absence of peculiar
velocities, an ellipsoid reduces smoothly to the usual real space
spherical linking volume, but tests on mock 2dFGRSs reveal that a cylinder
is slightly more effective at recovering groups that trace the
underlying dark matter haloes. This empirical motivation leads to a
cylindrical linking volume being employed throughout this paper.

\subsection{Empirically motivated fine tuning}\label{ssec:mods}

While the picture painted in section~\ref{ssec:link} is pleasingly
simple, with both the aspect ratio
and the linking volume being independent of group mass, in practice
the optimum linking volume to recover the $b=0.2$ set of dark matter
groups from the galaxy survey is not quite this universal. There are a
number of reasons why this similarity breaks down. For instance,
the mass to light ratio in the mock catalogues varies with halo mass
such that light is less concentrated in more massive clusters. Also,
halo concentration depends on mass, and concentration affects the halo
velocity structure and total enclosed overdensity (when the halo edge
is defined by an isodensity contour). Using the galaxy population which
sparsely samples the dark matter groups, these factors combine to yield
a small systematic bias in the recovered group properties. Tests with
mock catalogues have shown that this produces
either small haloes with overestimated sizes (and velocity dispersions),
or large clusters with sizes that are
underestimated. Across the range of masses from $10^{13}-10^{15}\sMsol$ 
this amounts to a $20$ per cent effect in the projected size
and slightly smaller in the velocity dispersion. To correct for this
requires knowledge of the halo mass in which each galaxy resides. An
estimate of this quantity is made by measuring the galaxy density
relative to the background
in a cylinder with aspect ratio $R_{\rm gal}$ and a comoving projected size of
$1.5\sMpc$. This density contrast, $\Delta$, is then used to scale both the
size and aspect ratio of the linking cylinder according to
\begin{equation}
b=b_{\rm gal} \left(\frac{\Delta}{\Delta_{\rm fit}}\right)^{\epsilon_b}
\label{bfiddle}
\end{equation}
and
\begin{equation}
R=R_{\rm gal} \left(\frac{\Delta}{\Delta_{\rm fit}}\right)^{\epsilon_R},
\label{Rfiddle}
\end{equation}
where $\Delta_{\rm fit}$, $\epsilon_b$ and $\epsilon_R$ are parameters to
be fitted from the mock catalogues. This enables the removal of the
biases described above, while effectively increasing the spread in linking
lengths at a given redshift by a few tens of per cent.

\subsection{The mean galaxy number density}\label{ssec:nofz}

In addition to varying with redshift, the mean observed galaxy number
density, $n$, depends on the depth to which a particular region of sky
was surveyed, and the efficiency with which redshifts were measured.
The production of maps that describe the angular variation of the
survey magnitude limit, $\bjlim(\btheta)$, the redshift completeness,
$R(\btheta)$, and a function, $\mu(\btheta)$, related to the apparent
magnitude dependence of the redshift completeness, is described in
section~8 of Colless \etal (2001). These quantities, along with a
galaxy weight, $w$, that models the local completeness of the 2dFGRS,
were combined to define $n$ at the position of each galaxy.

Galaxy weights were calculated by removing all fields in the 2dFGRS
that have a completeness less than 70\% and then all sectors (areas
defined by the overlap of 2dFGRS fields) that have a completeness less
than 50\%. Rejecting all galaxies from fields of low completeness
eliminates from the sample the small amount of data that
was taken in poor observing conditions. Rejecting sectors
with low completeness removes regions which are incomplete due to the
fact that some
2dFGRS fields were not observed or are excluded by the above
cut. Unit weight is then assigned to all the galaxies of the parent
APM catalogue in the remaining sectors. All of these galaxies without
measured redshifts have their weight redistributed equally to
their 10 nearest neighbours with measured redshifts.
Since low completeness sectors are excluded, the weights produced are
never large.
Their mean value is $1.2$ with an rms dispersion of $0.2$.
The inverse of the weight, $1/w$, is a local measure of the
completeness in the 2dFGRS around each galaxy.

The model for the redshift completeness as a function of
apparent magnitude is
\begin{equation}
c_z(\bj,\btheta) = \gamma [ 1 -\exp(\bj-\mu(\btheta))],
\label{redcomp}
\end{equation}
where $\gamma$ is a normalization factor determined by the
overall completeness in a given direction. The incompleteness
in a given direction is a result not only of the failure
to obtain redshifts for some of the faintest galaxies,
but also of the failure to target galaxies either because the
corresponding 2dFGRS field was not observed
or because of constraints on the fibre positioning.
To determine the completeness assumed in constructing the mock
2dFGRS catalogues, $\gamma$ is set in each sector
by demanding that the completeness averaged over apparent magnitude
\begin{equation}
\bar c_z(\btheta) = \int_{\bjbright}^{\bjlim(\btheta)}
N(\bj) c_z(\bj,\btheta) d\bj / \int_{\bjbright}^{\bjlim(\btheta)}
N(\bj) d\bj
\label{czbar}
\end{equation}
be equal to the measured overall completeness, $R(\btheta)$,
in that sector. Here, the integrals are over the apparent magnitude
range of the survey, from a global bright magnitude limit to the
local faint magnitude limit. A simple power-law fit to the observed
number counts, $N(\bj)\propto 10^{0.5\bj}$, is used.
In contrast, to model the completeness in the genuine or
constructed mock catalogue, $\gamma$ is fixed at the position of each galaxy by
demanding that $\bar c_z(\btheta) = 1/w$. That is, the inverse of the
weight assigned to each galaxy is taken as a local measure of the
completeness in that direction.
In the case of the genuine survey this has the advantage
that it will automatically take account of any small
scale variation in the completeness that might occur
due to the constraints on fibre positioning.
Having fixed $\gamma$, the comoving number density of
galaxies at each angular position and redshift is computed
from the 2dFGRS luminosity function as
\begin{equation}
n(z,\btheta) = \int_{\bjbright}^{\bjlim(\btheta)} \Phi(M(\bj,z))
c_z(\bj,\btheta) d\bj.
\label{nofz}
\end{equation}
The luminosity function used here is that estimated by Norberg \etal
(2002), convolved with their model of the 2dFGRS magnitude measurement
errors.

\subsection{The linking criteria}\label{ssec:defn}

Taking all of these survey characteristics into account, and defining a maximum
perpendicular linking length in physical coordinates as $L_{\perp,{\rm
max}}$, the comoving
linking lengths associated with a particular galaxy are
\begin{equation}
\ell_\perp={\rm min} \left[L_{\perp,{\rm
max}}(1+z),\frac{b}{n^{1/3}}\right];~~ \ell_{||} = R \ell_\perp ~,
\label{ldefs}
\end{equation}
and two galaxies $i$ and $j$, at comoving distances $d_{c,i}$ and
$d_{c,j}$ with an angular separation $\theta_{ij}$, are linked together if
\begin{equation}
\theta_{ij} \le \frac{1}{2}\left(\frac{\ell_{\perp,i}}{d_{c,i}}+\frac{\ell_{\perp,j}}{d_{c,j}}\right)
\label{linkang}
\end{equation}
and
\begin{equation}
|d_{c,i}-d_{c,j}| \le \frac{\ell_{||,i}+\ell_{||,j}}{2}~.
\label{linkz}
\end{equation}
The conversion of observed redshift to comoving distance ($R_0 \chi$)
requires an assumption about the cosmological model. This
impacts both on the inferred galaxy comoving space density, as $n \propto
(\d V/\d z)^{-1} \propto [R_0S(\chi)]^{2/3}/H(z)^{1/3}$ (see \eg
Peacock 1999) and on the comoving distance for a given redshift. 
The angle subtended by the chosen linking length
scales like $\theta \propto n^{-1/3}/(R_0 \chi)$. Throughout
this paper $\Omega_0=0.3$ and $\Lambda_0=0.7$ will be used. This is
appropriate for the mock catalogues and not obviously very wrong for
the real one (Spergel \etal 2003). 
At the median survey redshift of $0.11$, this model
yields an angular linking length that is almost two per cent larger
(and a comoving distance that is about five per cent larger)
than an Einstein-de Sitter model.

\section{Constructing a mock 2\d FGRS}\label{sec:mock}

It is clearly important to understand how the galaxy groups discovered
by this percolation algorithm are related to the underlying
distribution of dark matter in the universe.  To address this issue
and, at the same time, determine the optimum set of parameters to use
in the group-finder, mock 2dFGRSs have been constructed using high
resolution N-body simulations of cosmological volumes and a
semi-analytical model of galaxy formation.

The main N-body simulation used is the $\Lambda$CDM GIF volume
described by Jenkins \etal (1998). The
density parameter is $\Omega_{0}=0.3$, the cosmological constant is
$\Lambda_{0}=0.7$ and the normalisation of density fluctuations is set
so that the present-day linear theory rms amplitude of mass
fluctuations in spheres of radius $8\sMpc$, $\sigma_{8}=0.9$. The box
size is $141.3\sMpc$. Another simulation with $288^3$ particles in
a $\Lambda$CDM box of length $154\sMpc$ with $\sigma_{8}=0.71$ has also
been used to test the sensitivity of the optimum group-finding
parameters to the amplitude of the mass fluctuations. 
It turns out that the optimum
parameter choice is insensitive to this change in $\sigma_8$, although
the amount of spurious contamination does increase by $\sim 10$ per
cent when this decrease in the contrast between groups and not groups
is applied. The GIF volume will be used in the
subsequent analysis.

Dark matter haloes are identified in these simulation cubes
using a friends-of-friends algorithm
with a linking length of $b=0.2$ times the mean interparticle simulation.
The kinetic and potential energies of grouped particles are computed and
unbound particles are removed from the group. Bound groups of 10 or more
particles are retained, giving a halo mass resolution of
$1.4\times 10^{11}\sMsol$ (see Benson \etal 2001
for further details).

The reference semi-analytical galaxy formation model of Cole \etal
(2000) is used to populate the bound haloes identified in the $z=0$
output of the N-body simulation following the prescription outlined in
Benson \etal (2000).  The halo mass resolution of the N-body
simulation in turn imposes a resolution limit in the semi-analytical
calculation. This corresponds to the absolute magnitude of central
galaxies that occupy dark matter haloes that have ten or more
particles. Scatter in the formation histories of galaxies and variable
amounts of dust extinction cause  a spread in the
relationship between the luminosity of a central galaxy and the mass
of the host halo.  As a working definition, the magnitude limit of a
$z=0$ volume-limited galaxy catalogue constructed from the N-body
simulation is taken to be $M_{b_{\rm J}}-5\log_{10}h =-17.5$; at
this luminosity, $90\%$ of central galaxies predicted in a
semi-analytical model calculation without any halo mass limitations
reside in haloes resolved by the simulation.

The luminosity resolution of the volume-limited catalogue, when
combined with the global $k+e$ correction adopted by Norberg \etal
(2002), implies that a mock 2dFGRS survey constructed from this
simulation output will only be complete above a redshift of
$z=0.08$. The median redshift of the 2dFGRS is $z=0.11$, so it is
desirable to extend the mock catalogue to fainter luminosities. This
was done by constructing a volume limited sample of galaxies from a
separate semi-analytical calculation for haloes with masses less than
the N-body resolution limit. These galaxies were then assigned at
random to the particles in the simulation that were not part of a
bound halo. This should be a reasonable approximation because the
clustering of dark matter haloes is almost independent of mass for
masses beneath the resolution limit of the GIF simulation (Jing 1998).
Using this technique, the luminosity limit was shifted to $M_{b_{\rm
J}}-5\log_{10}h =-16.0$, so that a flux limited catalogue
constructed from this output would be complete above a redshift of $z=0.04$.

A mock 2dFGRS was constructed from the volume-limited galaxy catalogue
by applying the following steps: \\ o) A monotonic transformation was
applied to the magnitudes given by the semi-analytical model,
perturbing them slightly so as to reproduce the 2dFGRS luminosity
function. The reason for doing this is that the galaxy luminosity
function of the semi-analytical model is not a perfect match to the
measured 2dFGRS luminosity function (see Fig.~1 of Benson \etal 2000)
and it is desirable that the selection function of the mock catalogue
should accurately match that of the genuine survey.  The magnitudes
are then perturbed using the model of the 2dFGRS magnitude measurement
errors described in Norberg \etal (2002). \\ i) The volume-limited
catalogue was replicated, about a randomly located observer, to take
into account the much greater depth of the 2dFGRS volume compared to
the size of the N-body simulation box. This has no impact upon the
tests of the group-finder presented here, although it does mean that
the mock is unsuitable for studying the clustering of
groups on scales approaching the size of the simulation box.\\ 
ii) Galaxies were then selected from within this volume by
applying the geometric and apparent magnitude limits of the 2dFGRS
survey defined by the map, $\bjlim(\btheta)$, of the survey magnitude
limit (see Section~2 and Colless \etal 2001 section~8). This produces
a mock catalogue in which every galaxy has a redshift.\\ iii) The
appropriate redshift completeness was then imposed sector by sector on
the mock catalogue, by randomly retaining galaxies so as to satisfy
the function $c_z(\bj,\btheta)$ (equation~\ref{redcomp}).\\ This
method ignores any systematic variation of the completeness within a
sector, and in particular it does not take account of close-pair
incompleteness on angular scales less than $\sim 0.75$ arcmin. 
To investigate the effect of this, a second set of
mocks was also produced, in which close pairs were identified in the
parent mock catalogue and preferentially rejected when reproducing the
incompleteness in each sector. Not all close pairs are missed in the
2dFGRS because the large overlaps between 2dFGRS fields permits different
members of close pairs to be targetted on different fields. One
statistic that can be used to quantify the level of close-pair
incompleteness is the ratio of the angular correlation function of
objects that have measured redshifts to that of the full parent
catalogue (Hawkins \etal 2003). 
To reproduce the appropriate level of close pair
incompleteness in the mock catalogue, the parameters of the rejection
algorithm (the angular scale of the close pairs and the fraction that
are rejected) were tuned so as to reproduce this statistic. In
practice, the recovered groups hardly varied at all with the inclusion
of the close pair incompleteness. This is because of the combination
of the large amount of field overlap in the 2dFGRS and the small
angles over which fibre clashes become important.

Every galaxy in the mock survey is either spawned by
a dark matter halo containing at least ten particles or is a
background galaxy that is located at the position of an ungrouped dark
matter particle. In the following section, the phrase {\it the number of
galaxies spawned by a particular dark matter halo} will be used to
refer to the subset of galaxies belonging to that halo
which make it through the observing procedure and into the mock survey.

The use of a realistic mock 2dFGRS catalogue to set the parameters
required for the group-finder and to interpret the nature of the derived
groups represents a clear advance over previous work,
which either neglected to present any such tests or, at best, used
mock catalogues constructed from dark matter only simulations.
One criticism that could be levelled at the approach presented here
is that it is model dependent.
However, direct tests have been performed to confirm that the optimum
group-finding parameters are insensitive to catalogues created with a
$20$ per cent lower value of $\sigma_8$ (and appropriate adjustments
to the semi-analytical galaxy formation model).
It should be borne in mind that the default model does
provide a very reasonable description of the real universe.
In particular, it is in excellent agreement with a
number of observables that have a direct bearing on group-finding in the
2dFGRS: the local luminosity function in the $\bj$-band (see figure 10 of
Madgwick \etal 2002; Cole \etal 2000), the clustering of luminous galaxies
(Benson \etal 2000) and the dependence of galaxy clustering on luminosity
(Benson \etal 2001; Norberg \etal 2001). Any alternative model would also
need to reproduce all of these relevant observations for the testing of
the group-finder to be similarly appropriate.

\section{Testing the group-finder}\label{sec:test}

The mock surveys, complete with their parent dark matter haloes,
provide a database on which to optimise the parameters of the
group-finder. These are mainly the maximum perpendicular linking length
($L_{\perp,{\rm max}}$), the aspect ratio of the linking cylinder
($R_{\rm gal}$) and the number of mean intergalaxy separations
defining the perpendicular linking length ($b_{\rm gal}$). In the remainder of
this paper, the additional tweaks
to the linking volume described in Section~\ref{sec:fof} will be set
to values obtained empirically:
\begin{equation}
\Delta_{\rm fit}=5;\quad \epsilon_b=0.04;\quad \epsilon_R=0.16.
\label{tweaks}
\end{equation}
The process of
deciding which is the best set of group-finding parameters requires a
definition of what is good. A good group-finder should find a high
fraction of the available groups, provide accurate estimates of their
size and velocity dispersion, avoid splitting them up into
subgroups and minimise the number of interloping galaxies. In an
attempt to quantify these qualities, the following four statistics have been
defined:\\
1. Completeness, $c$, is the fraction of detectable dark matter haloes
that have more than half of their spawned galaxies in a single galaxy
group. A dark matter halo is {\it detectable} if it spawns at least two
galaxies into the mock survey.\\
2. The median accuracy, $a_\sigma$ or $a_r$, is the median $\log_{10}$
of the ratio
of associated galaxy group to dark matter group velocity dispersion or
projected size. The {\it associated galaxy group}
is the one that both contains the largest number of the galaxies spawned by
this dark matter halo (or the group with most members overall if there
is a tie), and is associated to this halo. A galaxy group is
{\it associated} with the dark matter halo that spawned most of its
members. If this is not unique, then the most massive of the possible
dark matter haloes is chosen. With these definitions, it is possible
for more than one galaxy group to be associated with the same dark
matter halo, but that dark matter halo will have only one associated
galaxy group. The set of values from which $a_\sigma$
and $a_r$ come is constructed by matching every detected dark matter
halo with its associated galaxy group. A {\it detected dark matter halo} is
one with an associated galaxy group. The spread about the median
is described by the semi-interquartile range, $s_\sigma$ or
$s_r$.\\
3. Fragmentation, $f$, is the mean number of extra galaxy groups per
dark matter halo
having mass, defined later in this subsection, at least $0.2$ times
that of their associated dark matter halo for all detected dark matter
haloes.\\
4. The quality, $q$, of an individual halo to group
match is defined as
\begin{equation}
q=\frac{N_{\rm good}-N_{\rm bad}}{N_{\rm spawn}},
\label{q}
\end{equation}
where $N_{\rm good}$ is the number of member galaxies spawned by this
halo that are found in the associated galaxy group, $N_{\rm bad}$ is the
number of group members not spawned by this halo and
$N_{\rm spawn}$ is the total number of galaxies spawned by this dark
matter halo.\\
The velocity dispersions of the galaxy groups were calculated using a
variant of the gapper estimator described by Beers, Flynn \& Gebhardt
(1990). This is an efficient estimator that is resistant to outlying
velocities. Tests on groups found from mock catalogues showed that
this estimator gave an $a_\sigma$ that varied less with the minimum
number of galaxies per group than other choices. The gapper estimator
also yielded an $s_\sigma$ that was at least as low as the others for
groups of all sizes. In
detail this involves ordering the set of recession velocities $\{v_i\}$
of the $N$ group members and defining gaps as
\begin{equation}
g_i=v_{i+1}-v_i,~~~ i=1,2,...,N-1.
\end{equation}
Using the following weights
\begin{equation}
w_i=i(N-i),
\end{equation}
the gapper estimate of the velocity dispersion can then be written as
\begin{equation}
\sigma_{\rm gap}=\frac{\sqrt{\pi}}{N(N-1)}\sum_{i=1}^{N-1}w_i g_i.
\label{vgap}
\end{equation}
Under the assumption that one of the galaxies is moving at the
centre-of-mass velocity of the halo, which is certainly true for the
mock catalogues, the estimated velocity dispersion is multiplied by an
extra factor of $\sqrt{N/(N-1)}$ before the redshift measurement error,
$\sigma_{\rm err}$, is removed in quadrature, giving
\begin{equation}
\sigma=\sqrt{{\rm max}\left(0,\frac{N\sigma_{\rm gap}^2}{N-1}-\sigma_{\rm err}^2\right)}~~.
\label{sigma}
\end{equation}
Velocity dispersions of the parent haloes, which inevitably have many more
dark matter particles in them than there are galaxies in the
associated groups, were calculated as the root mean squared velocity
difference from the mean.

The projected size of groups and haloes, $r$, is defined as the weighted root
mean squared projected separation from the group centre of the
members. This calculation is performed taking into account the galaxy weights
resulting from the variable incompleteness, as described in 
Section~\ref{ssec:nofz}.
The centre is defined using an iterative method that first
calculates the arithmetic weighted mean position of the remaining galaxies then
rejects the most distant galaxy. When only two galaxies
remain, the position of the galaxy with a larger weight, or if these are the
same, the larger flux, is deemed to represent the
group centre. The projected group size is then the weighted rms projected
physical separation of the other $N-1$ galaxies from this central
galaxy.  This measurement, in conjunction with the velocity
dispersion, can be used to estimate the group mass as
\begin{equation}
M=A \frac{\sigma^2 r}{G},
\label{mdef}
\end{equation}
where the value of $A$ was chosen so that, for the optimum choice of
group-finding parameters described later in this section, the median
mass was unbiased. This led to a choice of $A=5.0$. The
fragmentation statistic was calculated using these galaxy group masses.

These definitions provide a framework within which comparisons can be
made between group catalogues returned by different group-finding parameters.
A good set of parameter values will yield a set of galaxy groups that
have a completeness near to $1$, a median accuracy of $0$ (this being the
log of the ratio of measured to true velocity dispersion or radius),
independent of the group mass or redshift, with a small spread about
this median, a fragmentation near to $0$, and a quality of $\sim 1$.

\subsection{Definitions}\label{ssec:defs}
\begin{figure}
\centering
\centerline{\epsfxsize=8.5cm \epsfbox{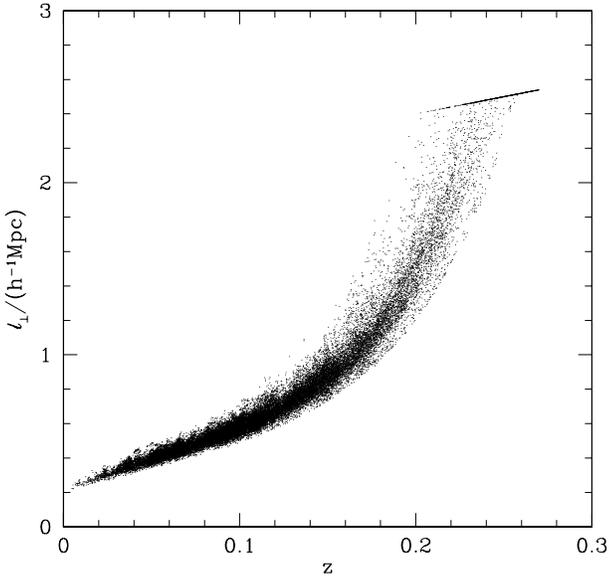}}
\caption{The variation of the comoving perpendicular linking length
with redshift when
$L_{\perp,{\rm max}}=2\sMpc$ and $b_{\rm gal}=0.13$ for each galaxy in
a mock SGP. Note
that the parameter $L_{\perp,{\rm max}}$ is in physical coordinates, whereas
$\ell_\perp$ is in comoving coordinates.}
\label{fig:bz}
\end{figure}

\begin{figure*}
\centering
\centerline{\epsfxsize=18.5cm \epsfbox{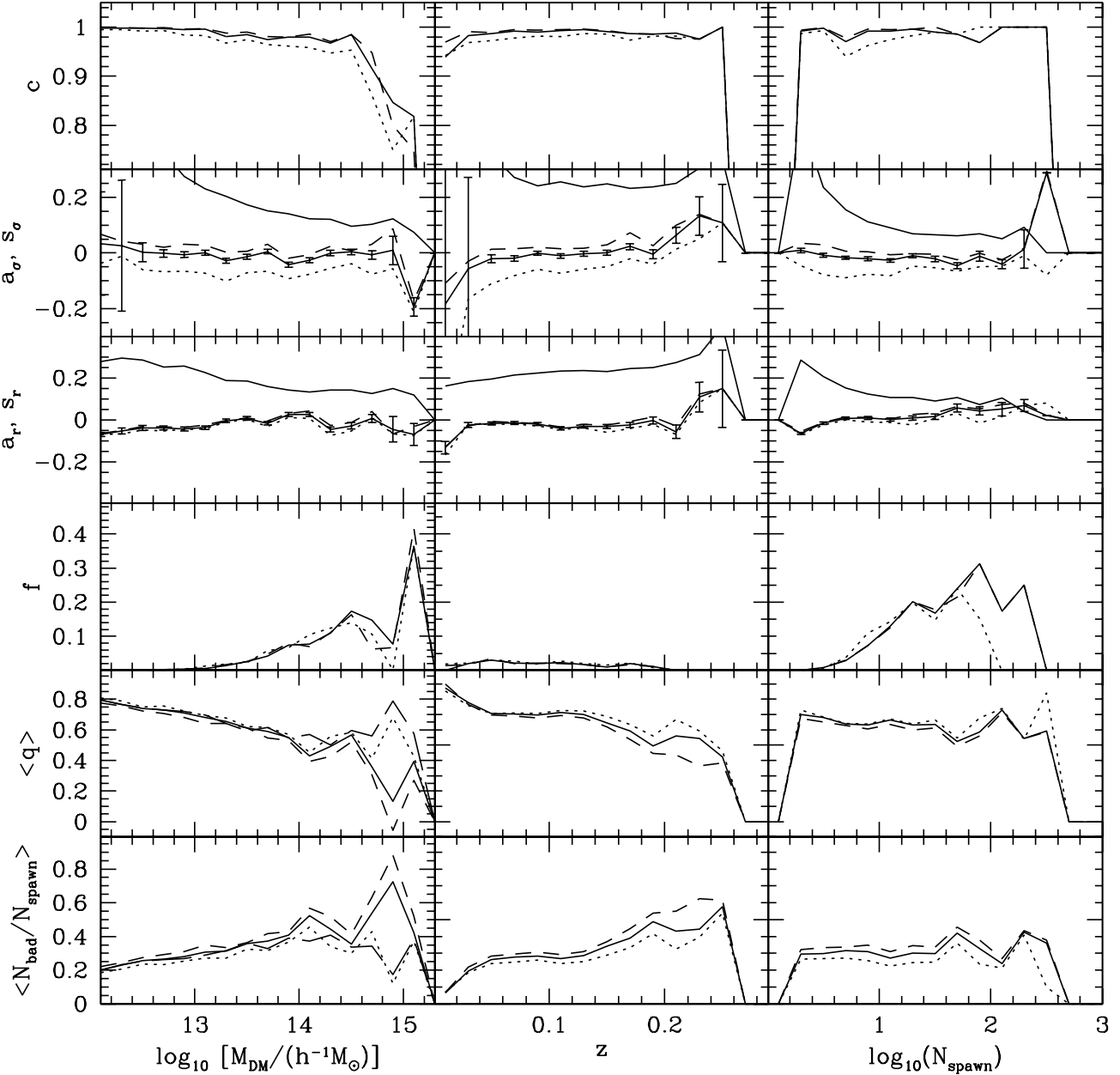}}
\caption{The effect of varying $R_{\rm gal}$ for an SGP mock
catalogue. $L_{\perp,{\rm max}}$ and $b_{\rm gal}$ are held fixed at
$2\sMpc$ and
$0.13$ respectively and $R_{\rm gal}$ is given values of $7$ (dotted),
$11$ (solid) and $15$ (dashed). The top row shows the
mean completeness as a function of dark halo mass, redshift and number of
galaxies spawned by the dark matter halo. The next rows show the median
accuracies of the velocity dispersions and projected sizes of the
galaxy groups, and the spread around the median value for the
$R_{\rm gal}=11$ case (upper solid lines without error bars).
The $1-\sigma$ errors shown on the median $R_{\rm gal}=11$ curves are
the errors on the mean accuracy
calculated from the spread, assuming that the individual accuracies are
distributed in a Gaussian fashion.
The mean number of additional galaxy groups
associated with detected dark matter haloes, as parametrised by the
fragmentation, is displayed in the next row, followed by
the mean quality of the group to halo matches in the penultimate row. 
This provides information about the difference between the numbers of good
and bad member galaxies. The mean number of bad interlopers relative to the 
number of galaxies spawned by the parent halo is shown in the final row. An
extra long-dashed curve is shown in the two lowest panels in the first column
to show how the quality of high mass halo matches improves when only $z<0.15$
groups are considered for the case of $R_{\rm gal}=11$. 
This also produces a corresponding decrease in the 
fraction of interlopers, as shown in the bottom panel.
}
\label{fig:varyR}
\end{figure*}
\begin{figure*}
\centering
\centerline{\epsfxsize=18.5cm \epsfbox{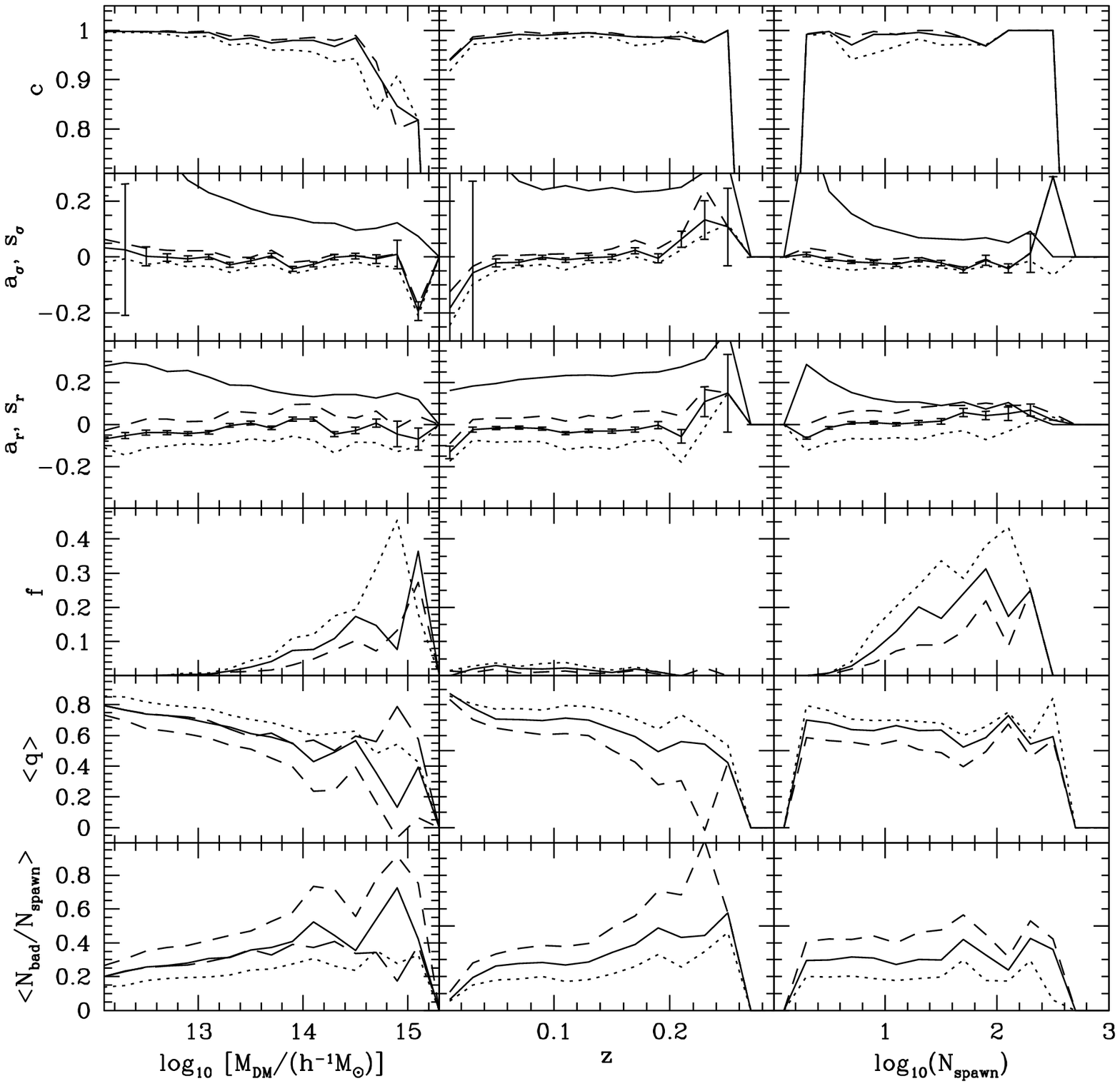}}
\caption{The effect of varying $b_{\rm gal}$ for an SGP mock catalogue. With
$L_{\perp,{\rm max}}=2\sMpc$ and $R_{\rm gal}=11$, $b_{\rm gal}$ was
given values of $0.11$ (dotted), $0.13$ (solid, and long-dashed for the 
$z<0.15$ subset in the two lowest panels in the first column) 
and $0.15$ (dashed). All
quantities shown are the same as those in Fig.~\ref{fig:varyR}.}
\label{fig:varyb}
\end{figure*}

\subsection{Parameter optimisation}\label{ssec:params}

In Section~\ref{sec:fof}, the three main free parameters of the group-finder
were described. These determine the overall size of the linking volume
through $b_{\rm gal}$, the maximum size of the linking volume via 
$L_{\perp,{\rm max}}$ and the aspect ratio of the linking volume, defined by
$R_{\rm gal}$. Applying the group-finder to the mock catalogues described in
Section~\ref{sec:mock} using various group-finding parameters, the
recovered galaxy groups can be compared with their parent dark matter
haloes, and the optimum set of parameters can be found. The following
three subsections detail the effect of varying each of these
parameters individually.

\subsubsection{Varying $L_{\perp,{\rm max}}$}\label{sssec:lpm}

The appropriate value for the maximum perpendicular linking length,
$L_{\perp,{\rm max}}$, should be similar to the size of the typical objects
that are detectable at larger redshifts, where the number density of
galaxies is low and this limit becomes relevant. Values around a
couple of physical $\Mpc$ are therefore a good place to
search. $R_{\rm gal}=11$ and $b_{\rm gal}=0.13$ have been fixed.
The justification for
choosing these particular parameter values is contained in the
following two subsections. Fig.~\ref{fig:bz} shows
the behaviour of the comoving perpendicular linking length with
redshift for $L_{\perp,{\rm max}}=2\sMpc$.
The increase of $\ell_\perp$ with redshift reflects the decreasing mean
observed number density of galaxies, and the spread in linking length
at a given redshift comes from the angular variation in survey depth and the
fraction of galaxies with measured redshifts. Larger values of
$L_{\perp,{\rm max}}$ lead to larger linking volumes at the highest
redshifts and some associated additional contamination. Decreasing
$L_{\perp,{\rm max}}$ yields underestimates of the projected sizes and
velocity dispersions of the groups at redshifts where the limit
affects the size of the linking volume. Thus
$L_{\perp,{\rm max}}=2\sMpc$ is chosen as a physically motivated compromise.


\subsubsection{Varying $R_{\rm gal}$}\label{sssec:lrat}

As was estimated in Section~\ref{sec:fof}, the appropriate choice for
the axis ratio of the linking cylinder is $\sim 10$. A variety of
values surrounding this one have been tried in conjunction with
$L_{\perp,{\rm max}}=2\sMpc$ and $b_{\rm gal}=0.13$.
Fig.~\ref{fig:varyR} compares the properties of the dark matter haloes
with those of their associated galaxy groups.  The projected group
sizes are fairly insensitive to $R_{\rm gal}$ as would be expected
considering that these changes only impact upon the line-of-sight
linking length. In contrast, the velocity dispersions are affected,
and the more elongated linking volumes yield larger values, with an
unbiased median being returned when $R_{\rm gal} \approx 11$. As the
linking volume is stretched, the completeness, fragmentation and
quality change only very slightly, with the most significant change
being a decrease of the quality of the group matches at higher
redshift. Note that the drop in quality for the most massive haloes is driven
by the large number of high redshift, poorly sampled groups. To illustrate 
this, an additional long-dashed 
line is included in the lowest two panels in the 
first column of Fig.~\ref{fig:varyR} showing the results using only the
groups at $z<0.15$.
This demonstrates that considering only the nearby, well sampled
massive groups yields a mean halo quality, and interloper content, that is
comparable with the lower mass haloes.
Using the requirement that the median velocity dispersion
be faithfully measured selects $R_{\rm gal}=11$ as the best
choice. This is comparable with that suggested by the rough
calculation leading to equation~\ref{lrat2}. Note that without the
additional halo mass dependent tweaks to the linking volume discussed
in Section~2, the values of which are given in Section~\ref{ssec:defs}, 
there would be a gradient of $\sim -0.05$ in all median
accuracy curves as a function of dark matter halo mass.

\subsubsection{Varying $b_{\rm gal}$}\label{sssec:b}

Keeping $L_{\perp,{\rm max}}=2\sMpc$ and $R_{\rm gal}=11$ fixed, the
value of $b_{\rm gal}$,
the number of mean intergalaxy separations defining the perpendicular
linking length, was varied from $0.11$ up to $0.15$, and the parent
$b=0.2$ dark matter haloes were again compared with the resulting
galaxy groups. Fig.~\ref{fig:varyb} shows the results.
As the linking lengths are increased, the recovered median group
velocity dispersions and projected sizes also increase as larger volumes are
grouped together. This is in contrast to the results when $R_{\rm gal}$ was
varied and only the velocity dispersions were much affected. These changes
are significant for all parent dark matter
halo masses. Along with these variations, the increasing linking lengths
increase completeness, reduce fragmentation and decrease the quality
of the matches. The least biased accuracies are produced when
$b_{\rm gal}=0.13$, and this is the value adopted for the rest of this
paper.
\begin{figure}
\centering
\centerline{\epsfxsize=8.5cm \epsfbox{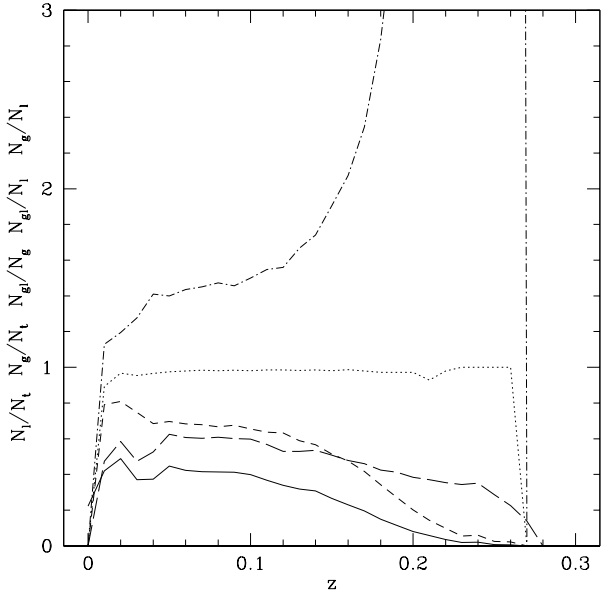}}
\caption{The redshift dependence of ratios of the
memberships of the three sets described in the text. The labels on the
y axis are ordered from bottom to top like the curves at
$z=0.1$. Thus, the fraction of all galaxies that have been spawned by
haloes that spawn at least two galaxies ($N_l/N_t$) is shown with the
solid line. The long dashed curve shows
the fraction of all galaxies that are put into groups ($N_g/N_t$).
The short dashed
curve traces the fraction of grouped galaxies that are in set $l$.
The dotted line shows the fraction of the $N_l$ galaxies that are
actually put into groups ($N_{gl}/N_l$),
and the dot-dashed curve displays the ratio of the number of
grouped galaxies to the number in set $l$.}
\label{fig:zdist}
\end{figure}

\subsection{Summary of group-finding parameters}\label{ssec:sum}

The results from the testing described in the previous three
subsections suggest that an appropriate choice of group-finding
parameters is $L_{\perp,{\rm max}}=2\sMpc$, $R_{\rm gal}=11$ and
$b_{\rm gal}=0.13$. These provide
a set of groups that have unbiased velocity dispersions and sizes. In
doing this, they contain almost all of the galaxies that should be
included in groups with at least $2$ members, and some interlopers as
well. Smaller values of all of the above three parameters should be
employed if reducing contamination is of greater importance than
capturing as many true group members while not overestimating the
velocity dispersion or projected size.
The level of contamination in the groups found by this particular
choice of group-finding parameters is illustrated in Fig.~\ref{fig:zdist}.
This shows the redshift dependence of ratios of the
memberships of the following three sets: 1) $t$, representing all
galaxies in the mock survey, 2) $g$, representing all grouped
galaxies and 3) $l$, representing galaxies that come from
dark matter haloes which have spawned at least two galaxies into the
mock survey, \ie those that could be linked to another galaxy spawned
by the same parent halo. $gl$ is used
to denote the overlap between sets $g$ and $l$, and $N_{i}$ is the
membership of group $i$.

The dotted curve in Fig.~\ref{fig:zdist} shows that the detected
groups contain almost all of the galaxies that belong in set $l$ for
all redshifts. For redshifts less than $0.1$, the fraction of all
galaxies that are grouped is $\sim 0.60$, as shown by the long dashed curve,
whereas the solid curve shows that the fraction of all galaxies in set $l$
is only $\sim 0.4$. Of the grouped galaxies in this redshift range,
only about $70$ per cent of them are members of set $l$. This quantity
is shown by the short dashed line in Fig.~\ref{fig:zdist}.

The accuracy statistic relates
only to the dark matter haloes that have spawned at least two galaxies
(\ie the detectable haloes), and have at least one associated galaxy group.
For the above choice of group-finding parameters, about $51$ per
cent of galaxy groups are the best matches to detectable dark matter
haloes, $\sim 47$ per cent are associated with dark matter haloes that
have only spawned one galaxy,
$\sim 2$ per cent are other matches with dark matter haloes and the
remaining $\sim 0.2$ per cent are made up completely from background
galaxies that have been spawned by no well defined dark matter halo.
Thus, a significant fraction of the recovered groups are not being used
to determine the group-finding parameters. These groups usually have
velocity dispersions and projected sizes that are larger than those of
their parent dark matter haloes. Consequently, the measured velocity
dispersion and projected size for typical groups
containing $2$ galaxies, will be overestimates of those of the underlying dark
matter halo. However, it would be inappropriate to adapt the
group-finding parameters to be unbiased when these spurious groups are
included, because this would inevitably spoil the results for the good
matches. 

\section{Application to the 2\d FGRS}\label{sec:2df}
\begin{figure*}
\centering
\centerline{\epsfxsize=21.5cm \epsfbox{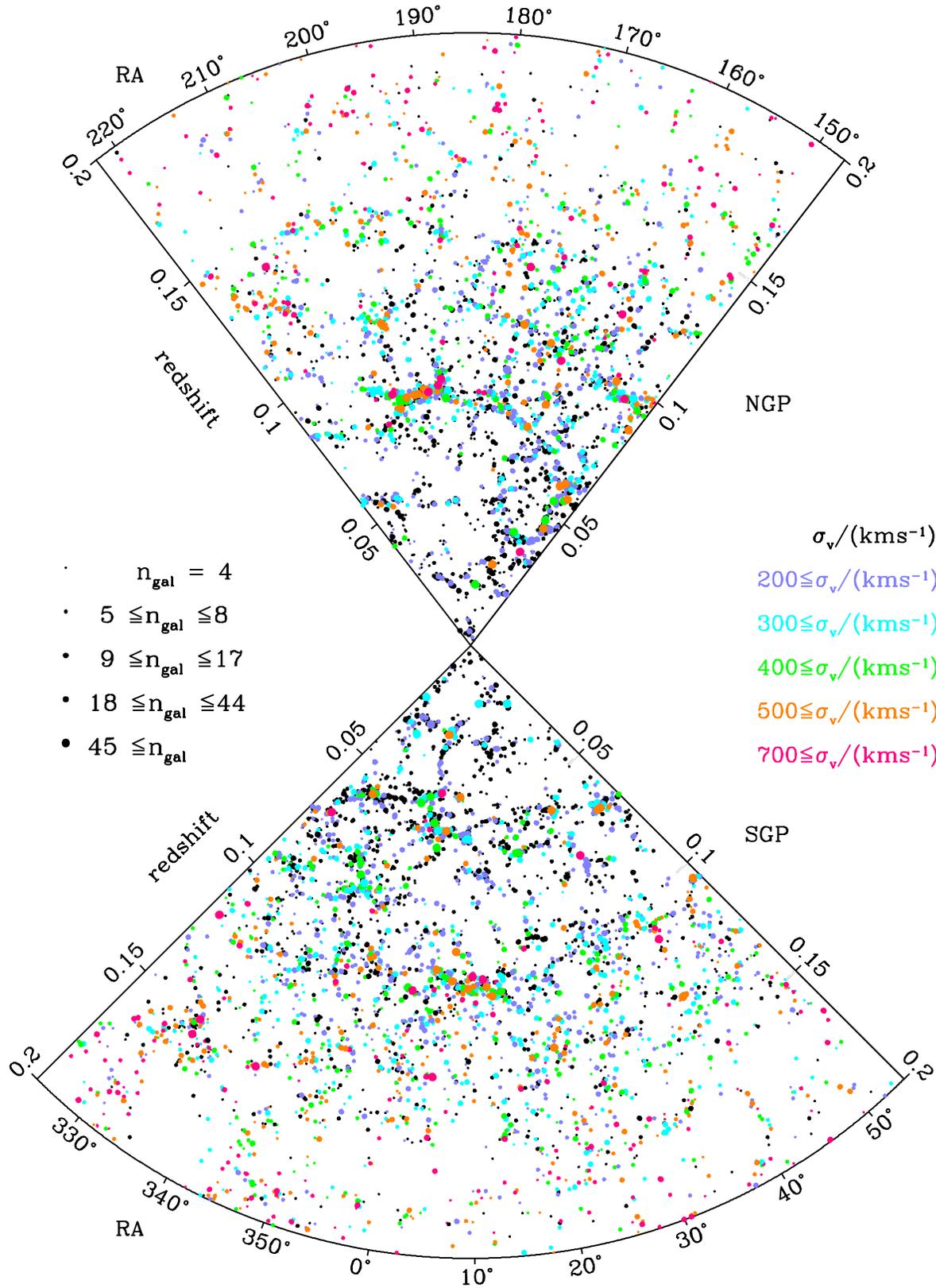}}
\caption{The spatial distribution of groups containing at least $4$
members in the NGP and SGP
regions. Dot colour and size represent the group velocity dispersion
and unweighted number of members respectively, as shown in the legend.}
\label{fig:dots}
\end{figure*}

\begin{figure*}
\centering
\centerline{\epsfxsize=15.5cm \epsfbox{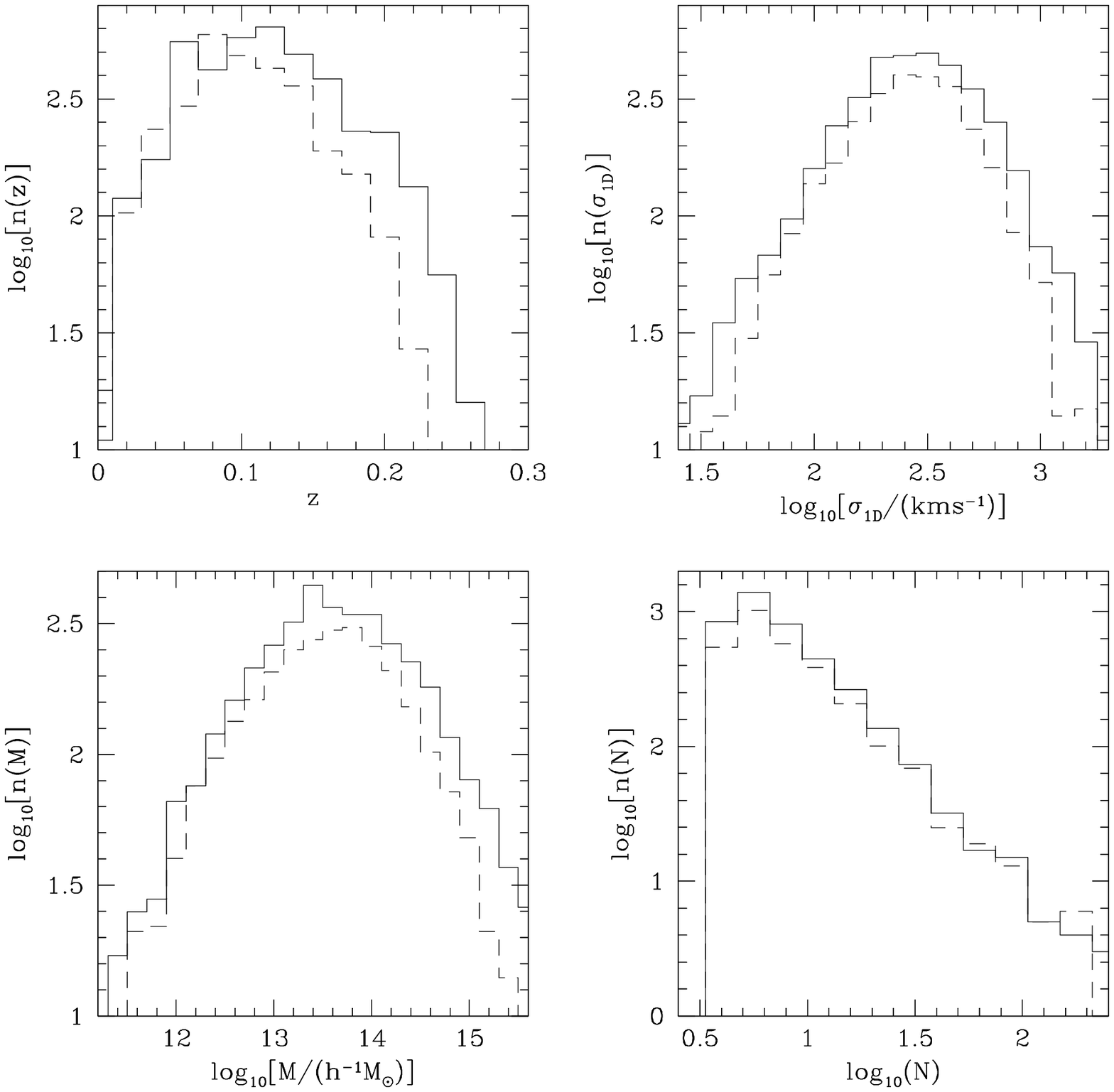}}
\caption{Histograms showing the distribution of group redshifts,
velocity dispersions, masses and the weighted number of members from the real
data. Only groups containing at least $4$ members have been included. 
The solid and dashed histograms show groups found in the SGP and
NGP respectively.}
\label{fig:props}
\end{figure*}

Together, the NGP and SGP regions in the 2dFGRS contain $191\,440$ galaxies
once cuts of 70\% and 50\% have been applied for field and sector
completeness respectively. This set of
galaxies, with their appropriate weights so that the mean
observed galaxy number density $n$ can be defined for each galaxy
according to equation~\ref{nofz}, has been used along with the
friends-of-friends group-finder described in Section~\ref{sec:fof} with
$L_{\perp,{\rm max}}=2\sMpc$, $R_{\rm gal}=11$ and $b_{\rm gal}=0.13$,
in order to find `real'
galaxy groups. The resulting catalogue places $55$ per cent of
all of the galaxies within $28877$ groups with at
least two members. A total of $7020$ groups are found with at least four
members, and their median redshift and velocity dispersion are $0.11$
and $260\skms$ respectively. The corresponding quantities for the sets of 
groups with at least 3 or 5 members are ($N_{\rm groups}$, median $z$, median 
$\sigma/\skms$) = (12566, 0.11, 227) and (4503, 0.11, 286).

The spatial distribution of the 2PIGGs containing at least $4$
galaxies is illustrated in Fig.~\ref{fig:dots}. Each dot represents a
group, with the colour and size representing the group velocity
dispersion and weighted galaxy content respectively. Typical dot sizes
decrease at large redshifts because the flux-limited survey means that
large distant groups are sampled with only a few bright galaxies. At
$z\gsim 0.15$ the typical velocity dispersion also grows. This happens
both because only the bigger groups have enough bright galaxies to be
recovered, and the smaller group memberships produce larger errors on
the measured velocity dispersions, thus increasing the abundance of
groups with apparently large velocity dispersions.
Figure~\ref{fig:props} shows the distributions of these groups with respect
to redshift, velocity dispersion, mass and weighted number of members,
for the NGP and SGP strips of the survey. While the SGP has almost
$50$ per cent more galaxies in it than the NGP, the fraction grouped
(0.56 and 0.54 for the NGP and SGP respectively) and the distributions
of properties of the resulting groups are very similar. The main
difference is in the redshift distribution of groups, where the lower
flux limit in the SGP betrays itself with a more extended distribution
than the NGP. Consequently, a few extra high velocity dispersion, or
equivalently high mass, clusters are found in the south. Although they
are not reproduced here, these distributions of group properties are
well matched to those found from the mock catalogues that were used to
calibrate the group-finder.

\begin{figure}
\centering
\centerline{\epsfxsize=8.5cm \epsfbox{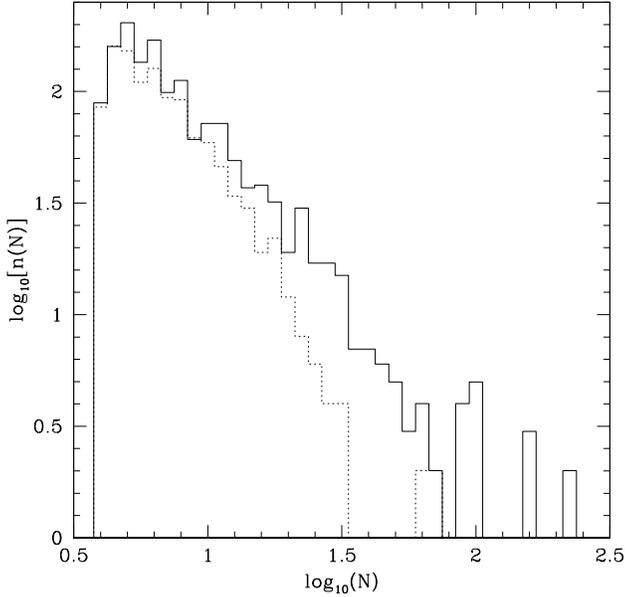}}
\caption{Histograms showing the distribution of weighted group
memberships for groups containing at least $4$ galaxies. These combine
the data from NGP and SGP, and include groups in the following two
redshift ranges: $0.04 \le z \le 0.08$ (solid) and $0.14 \le z \le
0.18$ (dotted).}
\label{fig:ndist}
\end{figure}

\begin{figure}
\centering
\centerline{\epsfxsize=8.5cm \epsfbox{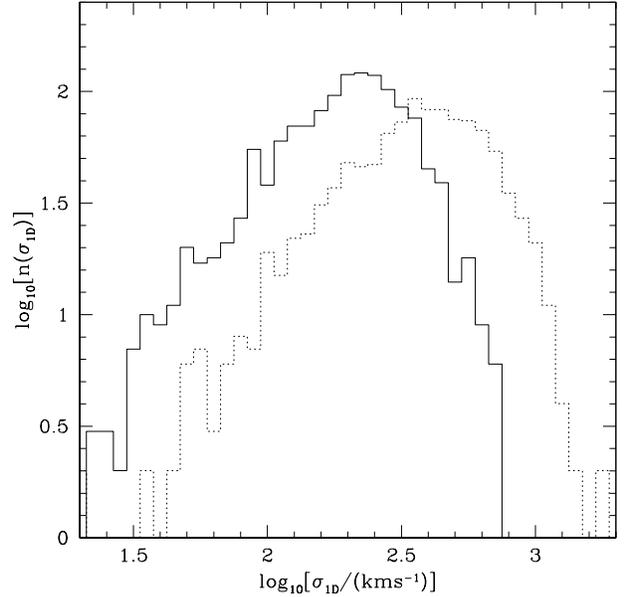}}
\caption{Histograms showing the distribution of velocity dispersions
for groups containing at least $4$ galaxies. These combine
the data from NGP and SGP, and include groups in the following two
redshift ranges: $0.04 \le z \le 0.08$ (solid) and $0.14 \le z \le
0.18$ (dotted).}
\label{fig:vdist}
\end{figure}

Figures~\ref{fig:ndist} and~\ref{fig:vdist} show how the group
memberships and velocity dispersions vary with redshift. This ties
together the information contained in Figures~\ref{fig:dots}
and~\ref{fig:props}, showing how the typical number of galaxies per
group decreases with increasing redshift, while the velocity dispersion
increases with increasing redshift. 

One interesting comparison can be made with the results
of Merch\'an \& Zandivarez (2002). They used the public 100k data release
version of the 2dFGRS and defined groups with a percolation algorithm
similar to the one used here. For 2PIGGs containing at least
four members, the mean redshift and velocity
dispersion, $0.11$ and $260\skms$, are very similar to their values
of $0.105$ and $261\skms$.
The total number of these groups has increased from the $2209$ that
Merch\'an \& Zandivarez found to $7020$ in the 2PIGG catalogue.
This factor is similar to the increase in the 
total numbers of galaxies being used, \ie roughly the same fraction of
galaxies are being grouped in both cases.

\section{Conclusions}\label{sec:conc}

A friends-of-friends percolation algorithm has been described,
calibrated and tested
using mock 2dFGRSs and then applied to the real 2dFGRS in order to
construct the 2dFGRS Percolation-Inferred Galaxy Group (2PIGG) catalogue.

From the mock catalogues, it is possible to determine the typical
accuracies with which velocity dispersions and projected halo sizes
are recovered. For detectable haloes recovered with galaxy groups
containing at least $4$ members the estimates of the velocity
dispersion are unbiased in the median and
the ratio of inferred to true velocity dispersion has a
semi-interquartile range of $\sim 30$ per cent.
The corresponding number for the
spread about the median ratio of inferred to true projected size is
$\sim 35$ per cent. Without the use of detailed mock catalogues to
calibrate the group-finder, the ability to derive science from the
2PIGG catalogue would be compromised by the uncertainty in how the
recovered groups related to the underlying distribution of
matter. This is thus a crucial step in the group-finding procedure.

When applied to the two contiguous patches in the real 2dFGRS,
containing $\sim 190\,000$ galaxies, this percolation algorithm groups $55$ per
cent of the galaxies into $\sim 29\,000$ groups containing at least two
members. Focusing on those groups with at least $4$ members, their
median redshift and velocity dispersion are $0.11$ and $260\skms$
repectively. More detailed distributions of fundamental group properties are
characterised in Section~\ref{sec:2df}.

This 2PIGG catalogue is the largest currently available set of groups. It
should provide a useful starting point for a number of studies
concerning large scale structure, galaxy group properties and the
environmental dependence of galaxy properties. 
Upcoming papers will describe in more detail the contents of the
groups, for instance the galaxy luminosity functions within groups, the
mass-to-light ratios of groups, the manner in which galaxies are
apportioned to different groups and the spatial distribution of galaxies
within groups. The spatial abundance of groups both as a function of
total group luminosity and mass will also be investigated.
The catalogue,
including basic group properties, is available on the WWW, at
http://www.mso.anu.edu.au/2dFGRS/Public/2PIGG/. A description of the
contents of this web page is given in the appendix.

\section*{ACKNOWLEDGMENTS}

VRE and CMB are Royal Society University Research Fellows. JAP is
grateful for a PPARC Senior Research Fellowship.

\begin{appendix}
\section[]{Details of the contents of the web page}\label{app:www}

\begin{table*}
\begin{center}
\caption{Data for a small subset of the galaxies in the NGP.
The quantities listed for each galaxy are:
(1) right ascension in radians (1950 coordinates);
(2) declination in radians (1950 coordinates);
(3) redshift;
(4) $\bj$ magnitude;
(5) limiting $\bj$ magnitude at this point in the survey;
(6) $n(z,\theta)$ at this point in the survey (see equation~\ref{nofz});
(7) galaxy weight, $w$, as described in Section~\ref{ssec:nofz};
(8) the group number to which each galaxy is assigned (zero is ungrouped).}
\begin{tabular}{llllllll} \hline
~~~(1) & ~~~~~(2) & ~~~(3) & ~~~(4) & ~~(5) & ~~~~~~~~~(6) & ~(7) & ~~~~~~(8)\\
~~~RA & ~~~~~~$\delta$ & ~~~~$z$ & ~~~$\bj$ & $b_{\rm J,lim}$ &
$n(z,\theta)/(\Mpc)^{-3}$ & ~~$w$ &group number \\
 \hline
2.66161&-0.06294&0.0976 & 18.379 & 19.394 &~~~~~0.017414 &1.00&\,~~~~~~~~~0\\
2.66142&-0.05574&0.0507 & 18.739 & 19.396 &~~~~~0.042268 &1.10&\,~~~~~~~~~0\\
2.66116&-0.05645&0.0770 & 18.665 & 19.395 &~~~~~0.026653 &1.00&~~~~~3018\\
2.66114&-0.05858&0.0582 & 18.345 & 19.396 &~~~~~0.039316 &1.00&~~~~~1711\\
2.66133&-0.06005&0.0602 & 19.205 & 19.394 &~~~~~0.037623 &1.00&~~~~~1711\\
2.66131&-0.06394&0.2049 & 19.052 & 19.394 &~~~~~0.001059 &1.10&\,~~~~~~~~~0\\
2.66090&-0.04524&0.0427 & 18.964 & 19.395 &~~~~~0.051024 &1.10&\,~~~~~~917\\
2.66086&-0.05390&0.0413 & 17.949 & 19.393 &~~~~~0.057687 &1.00&\,~~~~~~890\\
2.66092&-0.05668&0.0769 & 19.113 & 19.395 &~~~~~0.026625 &1.00&~~~~~3018\\
2.66087&-0.05679&0.0391 & 18.346 & 19.395 &~~~~~0.060466 &1.00&\,~~~~~~890\\
2.66099&-0.06091&0.0571 & 18.929 & 19.391 &~~~~~0.040111 &1.00&~~~~~1711\\
2.66109&-0.07300&0.2251 & 19.149 & 19.397 &~~~~~0.000540 &1.10&\,~~~~~~~~~0\\
2.66052&-0.05364&0.0778 & 18.483 & 19.392 &~~~~~0.026245 &1.00&\,~~~~~~~~~0\\
2.66054&-0.07209&0.1295 & 18.746 & 19.394 &~~~~~0.007955 &1.10&~~~~~7986\\
2.66038&-0.05148&0.0777 & 17.698 & 19.390 &~~~~~0.026254 &1.00&\,~~~~~~~~~0\\
\hline
\end{tabular}
\label{tab:gals}
\end{center}
\end{table*}

\begin{table*}
\begin{center}
\caption{Data for all 2PIGGs containing at least $100$
galaxies. The group quantities listed are:
(1) number of member galaxies;
(2) right ascension of the group centre in radians (1950
coordinates);
(3) declination of the group centre in radians (1950 coordinates);
(4) group redshift;
(5) rms projected galaxy separation in$\sMpc$;
(6) group velocity dispersion in$\skms$.}
\begin{tabular}{llllll} \hline
~(1) & ~~~(2) & ~~~~~(3) & ~~~(4) & ~~~~~~~(5) & ~~~~~~~(6) \\
$n_{\rm gal}$ & ~~~RA & ~~~~~~$\delta$ & ~~~~$z$ & rms $r/(\Mpc)$ &
 ~~$\sigma/(\kms)$ \\
 \hline
117 &  2.73888 & -0.05244 &  0.0346 & ~~~~~~1.12 & ~~~~~1069\\
120 &  2.86074 & ~0.03254 &  0.0397 & ~~~~~~1.04 & ~~~~~~437\\
121 &  3.40454 & -0.03863 &  0.0831 & ~~~~~~1.17 & ~~~~~~608\\
104 &  3.47684 & -0.01146 &  0.0836 & ~~~~~~1.11 & ~~~~~~668\\
158 &  3.38612 & -0.02542 &  0.0841 & ~~~~~~1.47 & ~~~~~~600\\
147 &  3.34815 & -0.02188 &  0.0856 & ~~~~~~1.64 & ~~~~~~778\\
143 &  3.38969 & -0.06855 &  0.0839 & ~~~~~~1.19 & ~~~~~~763\\
163 &  3.52656 & -0.02809 &  0.0858 & ~~~~~~1.48 & ~~~~~~786\\
112 &  2.96671 & ~0.01228 &  0.1021 & ~~~~~~2.38 & ~~~~~~566\\
140 &  0.10052 & -0.58160 &  0.0496 & ~~~~~~1.06 & ~~~~~~562\\
128 &  6.26000 & -0.61107 &  0.0490 & ~~~~~~1.10 & ~~~~~~702\\
159 &  5.86903 & -0.53810 &  0.0580 & ~~~~~~1.31 & ~~~~~~555\\
125 &  0.03848 & -0.50754 &  0.0611 & ~~~~~~0.98 & ~~~~~~654\\
119 &  0.82564 & -0.47254 &  0.0678 & ~~~~~~0.90 & ~~~~~~696\\
151 &  0.64879 & -0.58225 &  0.0773 & ~~~~~~1.66 & ~~~~~~762\\
116 &  0.06298 & -0.55218 &  0.1064 & ~~~~~~1.30 & ~~~~~~535\\
131 &  0.17340 & -0.50254 &  0.1081 & ~~~~~~1.86 & ~~~~~~722\\
199 &  0.16076 & -0.50887 &  0.1120 & ~~~~~~2.63 & ~~~~~~616\\
\hline
\end{tabular}
\label{tab:eg}
\end{center}
\end{table*}

http://www.mso.anu.edu.au/2dFGRS/Public/2PIGG/ contains: \\
1) the lists (NGP and SGP) of galaxies and the index of the groups in
which they are placed, \\
2) lists of group properties for the 2PIGGs, \\
3) the equivalent lists for some mock catalogues. \\
To illustrate the type of data that are available, 
Table~\ref{tab:gals} shows the information provided for a subset of
the galaxies and
Table~\ref{tab:eg} contains a list of all 2PIGGs containing at least
100 members.

\end{appendix}

\end{document}